\documentclass[preprint]{aastex63}
\usepackage{natbib, bm}

\newcommand{\pd}[2]{\displaystyle \frac{\partial #1}{\partial #2}}
\newcommand{\kak}[1]{\left( #1 \right)}

\shorttitle{Subsurface magma ocean on Io}

\shortauthors{Miyazaki and Stevenson}
\submitjournal{Planetary Science Journal}

\begin{document}
\title{A subsurface magma ocean on Io:\\ Exploring the steady-state of partially molten planetary bodies}

\correspondingauthor{Yoshinori Miyazaki}
\email{ymiya@caltech.edu}

\author[0000-0001-8325-8549]{Yoshinori Miyazaki}
\affil{Division of Geological and Planetary Sciences, California Institute of Technology\\
Pasadena, CA 91125 USA}

\author[0000-0001-9432-7159]{David J. Stevenson}
\affil{Division of Geological and Planetary Sciences, California Institute of Technology\\
Pasadena, CA 91125 USA}



\begin{abstract}
Intense tidal heating within Io produces active volcanism on the surface, and its internal structure has long been a subject of debate. A recent reanalysis of the Galileo magnetometer data suggested the presence of a high melt fraction layer with $>$50~km thickness in the subsurface region of Io. Whether this layer is a ``magmatic sponge'' with interconnected solid or a rheologically liquid ``magma ocean'' would alter the distribution of tidal heating and would also influence the interpretation of various observations. To this end, we explore the steady-state of a magmatic sponge and estimate the amount of internal heating necessary to sustain such a layer with high degree of melting. Our results show that the rate of tidal dissipation within Io is insufficient to sustain a partial melt layer of $\phi>0.2$ for a wide range of parameters, suggesting that such a layer would swiftly separate into two phases. Unless melt and/or solid viscosities are in the higher end of the estimated range, a magmatic sponge would be unstable, and thus a high melt fraction layer suggested in \cite{Khurana2011} is likely to be a subsurface magma ocean.
\end{abstract}

\section{Introduction}
Io, the innermost among four Galilean satellites, is tidally heated by forced eccentricity, exhibiting the most active surface volcanism in our Solar System. The absence of impact craters, the presence of substantial topography, and globally distributed volcanoes are best explained by the dominance of melt transport through a mostly rigid lid \citep{OReilly1981, Moore2003}. The extent of partial melting in the interior, however, remains elusive \citep[e.g.,][]{Bierson2016, Spencer2020a}. The internal structure characterizes the mode of tidal dissipation, including its magnitude, phase, and distribution, and thus is crucial for delineating the energy budget of a tidally heated body.

The reanalysis of the magnetometer data collected by the Galileo mission has shown a strongly induced magnetic signature within Io, and such an induction feature suggests the presence of a global high melt fraction ($>$20\%) layer thicker than 50~km in a subsurface region \citep{Khurana2011}. This has raised the question of whether this layer is a rheologically liquid ``magma ocean'' or a melt-rich ``magmatic sponge'' with interconnected silicate solid (Figure~\ref{fig_cartoon}a, b). The existence of a magma ocean would decouple the lithosphere from the interior, and in the original model for tidal heating proposed by \cite{Peale1979}, the tidal heating is then enhanced by the far larger strain (larger Love number) of the overlying shell. A magma ocean would likely not be internally very dissipative because of its low viscosity, while a magmatic sponge might be highly dissipative. We do not seek to resolve the dissipation issue here, but a magma ocean is likely to alter the internal distribution of tidal heating and may erase correlation between the depth of tidal dissipation and surface volcanism pattern. A concentration of volcanoes at low latitudes has been interpreted as a concentration of tidal dissipation in the shallow mantle \citep[e.g.,][]{Tackley2001, Tyler2015}, but a magma ocean could buffer the localization of magma upwelling, and the spatial distribution of eruptions could rather be a result of heterogeneity in lithospheric weakness.  

In this paper, we build a 1-D melt percolation model and examine the stability of a partially molten layer. A partial melt under large-scale deformation is known to be unstable and accumulates in veins \citep{Stevenson1989}, so melt migration in Io may not necessarily be characterized by percolation. Veins could eventually be interconnected, and melt may efficiently be drained to the surface. The growth of the instability, however, is yet to be quantified due to its nonlinearity, and how efficiently partial melt would escape to the surface remains unclear. Instead, we assume percolative flow and show that, even in this conservative approach, a mixture of interconnected melt and solid would swiftly separate into two phases. Our results indicate that a melt-rich layer beneath the Io's subsurface is thus more likely to be a magma ocean rather than a magmatic sponge. The observations of libration amplitude \citep{VanHoolst2020} or the Love number could potentially resolve the two scenarios. The latter will be measured by the Juno extended mission, perhaps to sufficient accuracy to delineate between a magma ocean and a magma sponge, and we aim to make a theoretical prediction before the planned measurements of the Love number. This paper is organized as follows. In Section 2, we describe our model set-up, focusing on the boundary conditions of the model. Model results with a range of parameter values are presented in Section 3, and the presence of a subsurface magma ocean is suggested in the last part of this paper.

\section{Methods} \label{sec:method}
Melt percolation within a tidally-heated mantle is considered in this study to examine the stability of a partially molten layer. Instead of solving for a time evolution, we explore steady-state structures and investigate the relation between the degree of tidal dissipation within the mantle and the resulting extent of partial melting. The goal is to estimate the amount of tidal heating necessary to maintain a ``magmatic sponge'' and to show that tidal dissipation within Io is insufficient to sustain interconnected solid containing a high degree of partial melting. For simplicity, a spherically symmetric structure is assumed, and 1-D radial profiles of percolation velocity and melt fraction are solved for various degrees of tidal dissipation. 

Given the uncertainty in the state of Io, this study aims to model heat transport with a minimum number of parameters. The parameters characterizing the mantle system are limited to solid viscosity, $\eta_s$, the ratio of melt viscosity to permeability, $\eta_l/k_m$, and melt-solid density difference, $\Delta \rho$ (300~kg~m$^{-3}$ in this study), \citep[e.g.,][]{Stevenson1991, Spiegelman1993a} and we explore how the profile of melt fraction changes with different rates of tidal-heating, $\psi$. In addition, we specify the liquid overpressure at the top boundary, $\tau_0$. A reasonable value for the liquid overpressure is discussed in detail in Section~\ref{sec:boundary}, but fortunately, the choice of overpressure does not affect our key results. We note that the top boundary does not necessarily represent the crust-mantle boundary. The crust and the partial melt region are likely to be separated by a transition zone of several kilometers in thickness, which is characterized by transient behaviors and may not be described by percolation or eruption. We discuss the nature of such a transition zone in Section~\ref{sec:boundary}. It is noted that partial melt is compositionally different from the bulk mantle, and thus the near-surface solid layer, which is consisted of a solidified partial melt, would also differ in composition from the mantle and can be defined as a crust.

The model presented here can be applied to any body size, but with an application to Io in mind, a planet with 1820 km radius and a partially molten mantle of $\sim$200--700 km is considered in this study. We first summarize the governing equations and then discuss boundary conditions.

\subsection{Governing equations}
Governing equations for melt percolation within a solid-melt mixture consist of the conservation of mass, 
\begin{equation} \label{cmass}
	\pd{\phi}{t} = - \nabla \cdot \kak{\phi v_l} + \Gamma,
\end{equation}
the conservation of momentum \citep{Stevenson1991},
\begin{equation} \label{eom}
	\kak{1 - \phi} \Delta \rho g = \frac{\eta_l}{k_m} \phi \kak{v_l - v_s} + \frac{d}{dz} \kak{ \kak{1 - \phi} \kak{\zeta + \frac{4}{3} \eta_s} \frac{dv_s}{dz} },
\end{equation}
and the conservation of energy \citep{Katz2008},
\begin{equation} \label{cene}
	\rho c_p \pd{T}{t} + \nabla \cdot \kak{(1-\phi) v_s \cdot \rho c_p T} +  \nabla \cdot \kak{\phi v_l \cdot \rho \kak{c_p T + L}} =  k \nabla^2 T + \psi.
\end{equation}
The parameters $\phi$ denotes volumetric melt fraction, $T$ is temperature, and $\Gamma$ is a change in melt fraction due to tidal heating. The constant $g$ indicates gravitational acceleration (1.5~m~s$^{-2}$), $\rho$ is the mantle density (3300~kg~m$^{-3}$), $c_p$ is specific heat capacity (1000~J~kg$^{-1}$~K$^{-1}$), $L$ is the latent heat of silicates ($4\times 10^5$~J~kg$^{-1}$), and $k$ is thermal conductivity (3.3~W~m$^{-1}$~K$^{-1}$). We set effective bulk viscosity, $\zeta$, to $\eta_s/\phi$ based on microscopic models \citep{Sleep1988, Hewitt2008}. 

By considering a horizontally uniform steady-state system, a frame-invariant variable $u \equiv \phi (v_l - v_s)$ can be defined \citep{Scott1984}, which becomes the same as the solid velocity under a barycentric coordinate. Using the separation flux $u$, Equations~(\ref{cmass}--\ref{cene}) are simplified to 
\begin{equation}  \label{eom2}
	\kak{1 - \phi} \Delta \rho g + \frac{d}{dz} \kak{\kak{1 - \phi} \kak{\zeta + \frac{4}{3} \eta_s} \frac{du}{dz}} = \frac{\eta_l u}{k_m},
\end{equation}
\begin{equation} \label{cene2}
  	\rho L \nabla \cdot \kak{ (1-\phi) u}  = k \nabla^2 T + \psi,
\end{equation}
and the non-dimensionalized version of the two equations are solved here (Appendix~\ref{sec:nond}). The latter equation indicates that heat loss by the advection of latent heat mostly balances tidal heating. Thermal conduction is mostly negligible in the partial melt layer except in the near surface region where the overlying cold crust affects the thermal structure (Section~\ref{sec:litho}).   

\begin{figure}[htbp]
\centering
\includegraphics[width=.95\linewidth]{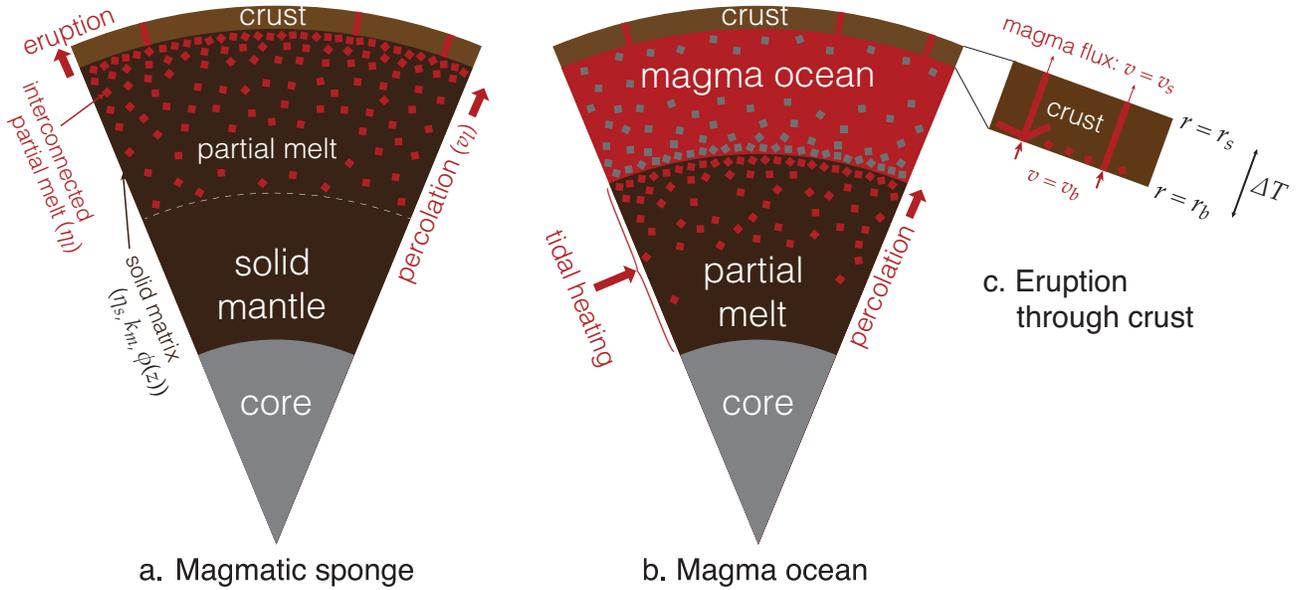}
\caption{Schematic illustration of Io's interior with (a) a magmatic sponge and (b) a subsurface magma ocean. An enlarged view of eruption through the crust is also shown on the right. Parameters defined in Sections~\ref{sec:method} and \ref{sec:litho} are labeled to the corresponding reservoirs. Our results suggest that a magmatic sponge would swiftly separate into two phases and create a magma ocean, which is a rheologically liquid layer with $\phi$$>$0.4. It is noted that the crust and the partially molten layer are connected by a transition zone, where upwelling magma solidifies to heat the subsiding crust (Section~\ref{sec:transit}).} 
\label{fig_cartoon}
\end{figure}

\subsection{Boundary conditions} \label{sec:boundary}
Boundary conditions of the momentum equation (Equation~\ref{eom}) pose a challenge in constraining the profile and advection of melt within the partially molten layer. Whereas the bottom boundary condition is straightforward ($\phi=0$, $u=0$), the top boundary involves interaction with the overlying layer. Percolated melt that reached the top of the partial melt layer has to be delivered to the planet's surface to release heat through eruption \citep{OReilly1981}. 

The top of the partially molten region is often considered to be a crust-mantle boundary, and the transition from a partial melt regime to a solidified crust has often been thought of as a ``boundary condition'' on the former. This, however, is a questionable approach since the two regimes are so different: the partial melt layer is dominated by viscous flow with pervasive yet microscopically distributed melt, whereas the crust is elastic-plastic with viscoelasticity in its lowermost regions, which is capable of temporal and spatial fluctuations that might include crack formation. In reality, the ``boundary'' is rather likely to be a transition zone of several kilometers in thickness (see Section~\ref{sec:litho}), and neither partial melt nor standard crack formation recipes may apply. The perplexing issue of correct petrology remains unsolved as well. 

While governing equations in the transition zone remain unresolved, energy balance requires that a large fraction of melt that originated in the partial melt region solidifies within the transition zone to heat the cold subsiding crust (see Section~\ref{sec:litho} for details). The upwelling melt flux is thus expected to decrease while going through the transition zone, whereas in most of the partially molten regions, the melt flux increases towards the surface to release tidal heating (Figures~\ref{fig_ref}-\ref{fig_solid}). This implies that $du/dz$ would be 0 in the vicinity of the boundary between the transition zone and the partial melt layer. Here, the transition zone is defined as a region where the standard percolation recipe does not apply, but it is unclear under what conditions percolative flow becomes inapplicable. 

Instead of determining the exact value of $du/dz$ at the boundary, we assume that the pressure difference between the liquid and solid phases is close to 0. The liquid overpressure is related to the compaction rate as 
\begin{equation} \label{diffP}
	\Delta P = P_{liq} - P_{sol} = \zeta \nabla \cdot v_s = -\zeta \frac{du}{dz} = \tau_0,
\end{equation}
where $z$-direction is defined positive towards the surface. The top boundary of the partial melt layer is set to have a small liquid overpressure of $\tau_0 \sim$1~MPa, which is typical stress needed to initiate fracture \citep{Sleep1988}. We, however, stress that standard crack formation may not apply, and $\tau_0$ can potentially take any value around 0~MPa. Values between -1 and 3~MPa are tested, but we find that changing $\tau_0$ by a few MPa has little influence on the overall results. Melt fraction at the top boundary, $\phi_{top}$, is set so that the liquid overpressure within the partial melt layer does not exceed $\tau_0$.

\subsection{Plausible range of rheological parameters} \label{sec:par}
Given the uncertainty in the Ionian interior, the expected range of $\eta_s$, $\eta_l$, and $k_m$ could span a few orders of magnitude. Solid viscosity $\eta_s$ is known to follow \citep{Hirth1996, Mei2000, Jain2019}
\begin{equation} \label{diff}
	\eta_s = \eta_0 \left(\frac{c_{w,0}}{c_w}\right) \left(\frac{d}{d_0}\right)^2 \exp \left( \frac{E}{R} \left( \frac{1}{T} - \frac{1}{T_0} \right) \right),
\end{equation}
where $c_w$ shows volatile content, $d$ is typical grain size, $E$ is the activation energy, $R$ is the universal gas constant, and the subscript 0 denotes a reference value. Diffusion creep, which is likely to be the dominant deformation mechanism under a near-solidus temperature, is assumed here, and Equation~(\ref{diff}) shows that solid viscosity decreases with a higher volatile content, a higher temperature, and smaller grain size. Sulfur-rich volcanism indicates that the interior of Io contains more volatile than the terrestrial mantle \citep[e.g.,][]{McGrath2000}, and basaltic lava composition implies that the mantle temperature is similar to Earth. Modeling of the eruption temperature also suggest a similar mantle temperature \citep{Keszthelyi2007}, but a lava temperature higher than $\sim$1800~K has been observed on some occasions \citep{Davies2001, deKleer2014}, possibly pointing to a hotter interior than Earth. Therefore, volatile content and interior temperature of Io suggest that the solid matrix of Io is less viscous than the terrestrial mantle. Grain size within Io remains uncertain because efficient grain growth may occur under a low-stress environment, but it is unlikely that the grain size becomes larger by an order of magnitude than in the Earth's interior ($\sim$1~cm). We thus estimate that solid viscosity would not be higher than 100~times the terrestrial value. Rock mechanics and the possibility of small-scale convection beneath oceanic lithosphere predict a value of $10^{19}$~Pa~s for Earth \citep{Davaille1994, Hirth1996, Dumoulin1999}, so we estimate that $\eta_s$ is smaller than $10^{21}$~Pa~s on Io. 

Melt viscosity follows a similar trend, and $\eta_l$ is smaller under a higher temperature and volatile content. The SiO$_2$ content also affects melt viscosity, with ultramafic magma having a significantly smaller viscosity than rhyolitic. Silicate lava on Io is best explained by basaltic or ultramafic compositions \citep{McEwen1998}, so we take the viscosity of basaltic melt as a reference value: $\sim$100~Pa~s under $\sim$1300~K \citep{Giordano2008}. With an abundant amount of sulfur in Io's magma and a potentially higher temperature, $\eta_l$ is expected to be smaller than the terrestrial value, thus $\eta_l<100$~Pa~s in Io's mantle. We note that this maximum value is subject to uncertainty because Io's interior and melt rheology could be controversial. Sulfur may be mainly stored in a segregated surface layer, and volatile-depleted melt in the interior may be highly viscous. Melt viscosity also increases when small crystals are abundantly suspended in the percolating melt phase. Under the same $\phi$, however, suspended crystals result in a more porous matrix and a higher permeability, so it could cancel out the effect of viscosity increase ($\eta_l/k_m$ in Equation~(\ref{eom2})). It is thus not straightforward to determine the upper bound of viscosity, and we may need to consider $\eta_l$$\ge$100~Pa~s as a higher-end value. The lowest possible values are not constrained here because, as shown later, a layer with a high melt fraction becomes increasingly unstable under lower melt or solid viscosities. The conclusion of our study thus remains unchanged even for small values of rheological parameters.

Permeability $k_m$ is also a function of grain size $d$, and dimension analysis suggests that $k_m$ is proportional to $d^2$. For a permeability model, a scaling relation $k_m = k_0 \phi^2$ is adopted based on a simple tube model. This is likely to be valid under a high melt fraction of around 0.1-0.2, which is the interest in this work. A permeability model for small porosity is subject to debate, which is applicable to studies of mid-ocean ridges, for example. Smaller values of the permeability prefactor $k_0$ and cubic dependence on melt fraction in the literature correspond to low melt fraction where the effects of non-uniform tubule radii and surface tension are pronounced. Such effect, however, is irrelevant to the melt fraction of interest here. The coefficient $k_0$ is thus approximated by $d^2/32$, and assuming a grain size of 1~mm to 1~cm, $k_0$ is in the range of $3 \times 10^{-8} - 3 \times 10^{-6}$. Large grain size may increase permeability, but it has the same effect as lowering $\eta_l$. A high-melt fraction layer would become increasingly unstable, and it would not change the conclusion of this study.

\section{Results} \label{sec:res}
We first consider a partially molten layer with a thickness of 400~km, corresponding to the upper half of the Ionian mantle, and different degrees of tidal heating are applied to this layer. For simplicity, tidal heating is uniformly distributed throughout the partial melt layer, but the volumetric heating rate may differ by a factor of $\sim$4 with depth \citep{Bierson2016}. the profile of melt fraction does not change significantly as long as the total amount of tidal heating remains the same. 

\begin{figure}[htbp]
\includegraphics[width=1\linewidth]{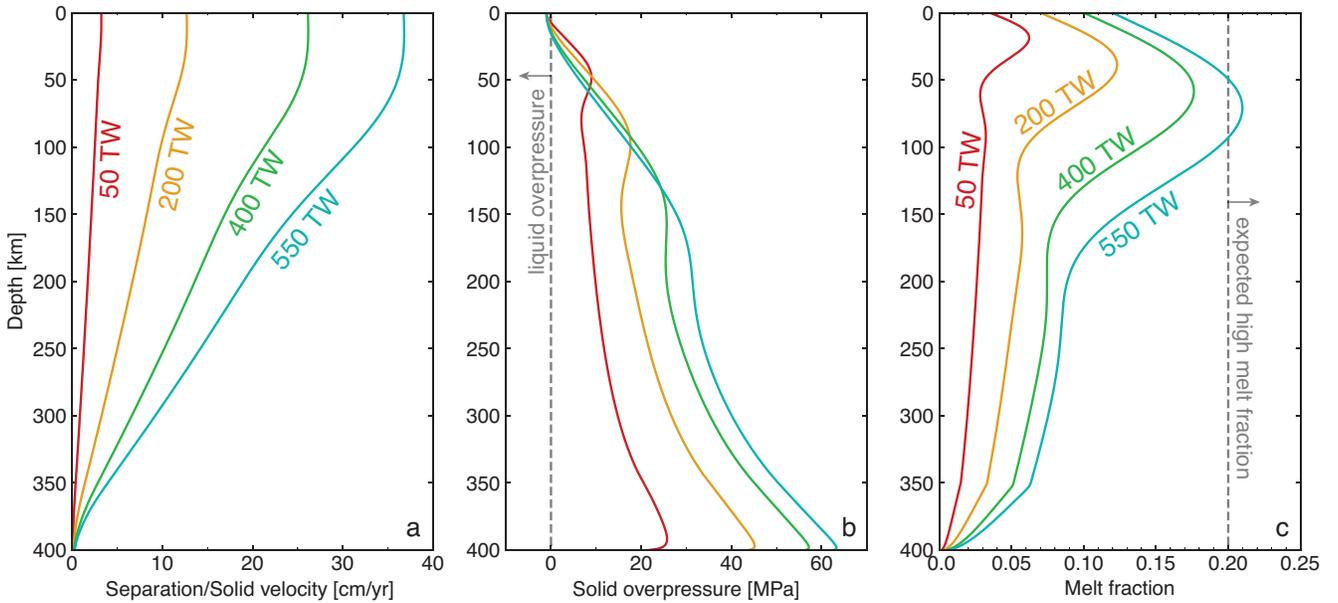}
\caption{Steady state solutions of (a) separation velocity (cm/yr), (b) solid overpressure (MPa), and (c) melt fraction for a case with $\eta_l = 10$ and $\eta_s = 10^{20}$~Pa~s. Colors indicate the amount of tidal heating, corresponding to 50 (red), 200 (orange),  400 (green), and 550 TW (blue). The thickness of a high melt fraction layer ($\phi>20\%$; gray dashed in (c)) exceeds 50~km only when the rate of tidal dissipation is larger than 550 TW.}
\label{fig_ref}
\end{figure}

Figure~\ref{fig_ref} shows a typical mantle structure under $\eta_l/k_0 = 10/2 \times 10^{-8}$ and $\eta_s = 10^{20}$~Pa~s. As expected, the degree of melting increases as tidal heating intensifies, and a larger separation flux is observed to allow for efficient heat loss. Whereas melt and solid velocities increase towards the surface, melt fraction exhibits an oscillatory behavior with a peak near the top of the partial melt layer \citep{Hewitt2008}. This peak arises due to a pressure difference imposed at the top boundary and can be understood from the net mass flux:
\begin{equation} \label{dcomp}
	\nabla \cdot \left( (1-\phi) u \right) \simeq - u \frac{d\phi}{dz} + (1-\phi) \frac{du}{dz} \ge 0.
\end{equation}
Tidal dissipation continuously melts the partially molten layer, so the net mass flux has to be positive at all depths to maintain a steady state. At the same time, cooling by the subsiding crust cools the uppermost region to create small or negative solid overpressure (Section~\ref{sec:boundary}), resulting in a negative value of the factor $d\phi/dz$. The melt fraction rapidly increases with depth in the subsurface region, and its value becomes higher by a factor of 2--4 compared to the average melt fraction. 

The thickness of a high melt fraction layer ($>$20~\%), however, only exceeds 50~km when tidal dissipation within the partially molten layer is larger than 550~TW (Figure~\ref{fig_ref}c). This is significantly larger than the estimated rate of tidal dissipation for Io, where astrometry \citep[93$\pm$19~TW;][]{Lainey2009} and ground-based observations of global thermal emission \citep[105$\pm$12~TW;][]{Veeder1994} both suggest a value of $\sim$100~TW. The latter may not reflect high-latitude heat flow and long-wavelength emission \citep{Stevenson1988}, and with the contribution from those sources, the global heat flow would increase to 122$\pm$40~TW \citep{Veeder2004} or potentially to a larger value. Yet, the magnitudes of heat loss and tidal dissipation are expected to be similar considering that Io has maintained thermal equilibrium in a long term, and the rate of tidal dissipation is likely to fall into a range of 100-200~TW. For parameters used in Figure~\ref{fig_ref}, tidal dissipation in Io is therefore insufficient to sustain a high-melt fraction layer, and a ``magmatic sponge'' is likely to swiftly separate into two phases. In the following sections, we explore whether such a shortage of tidal heating is ubiquitous among a plausible range of values. 
 
\subsection{Dependency on rheological parameters}
\begin{figure}[htbp]
\includegraphics[width=1\linewidth]{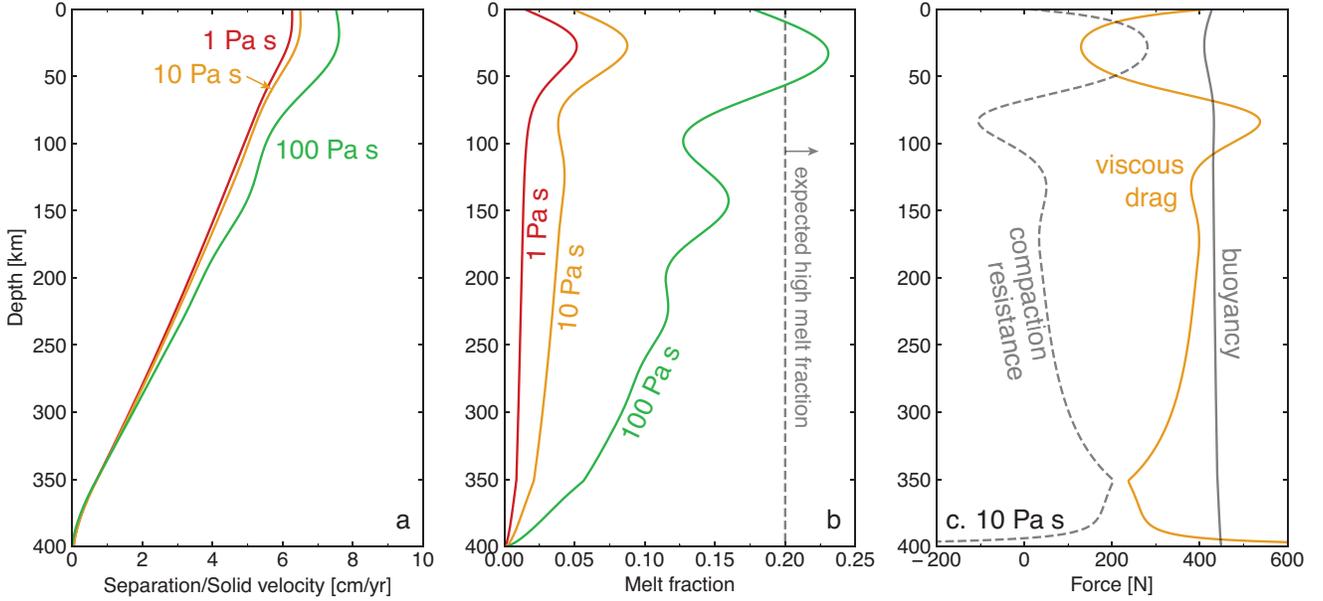}
\caption{(a, b) Steady state solutions of (a) separation velocity (cm/yr) and (b) melt fraction under $\eta_s = 10^{20}$~Pa~s and 100 TW of tidal dissipation within the layer. Colors indicate different melt viscosities, corresponding to 1 (red), 10 (orange), and 100~Pa~s (green). The thickness of a high melt fraction layer ($\phi>0.2$) exceeds 50~km only when $\eta_l$ is larger than 100~Pa~s. (c) Forces described in Equation~(2) for a case with $\eta_l = 10$~Pa~s. Buoyancy (gray solid) is plotted as positive upward, whereas viscous drag (orange) and the resistance of solid matrix to compaction (gray dashed) are positive downward.}
\label{fig_melt}
\end{figure}

\begin{figure}[htbp]
\centering
\includegraphics[width=.7\linewidth]{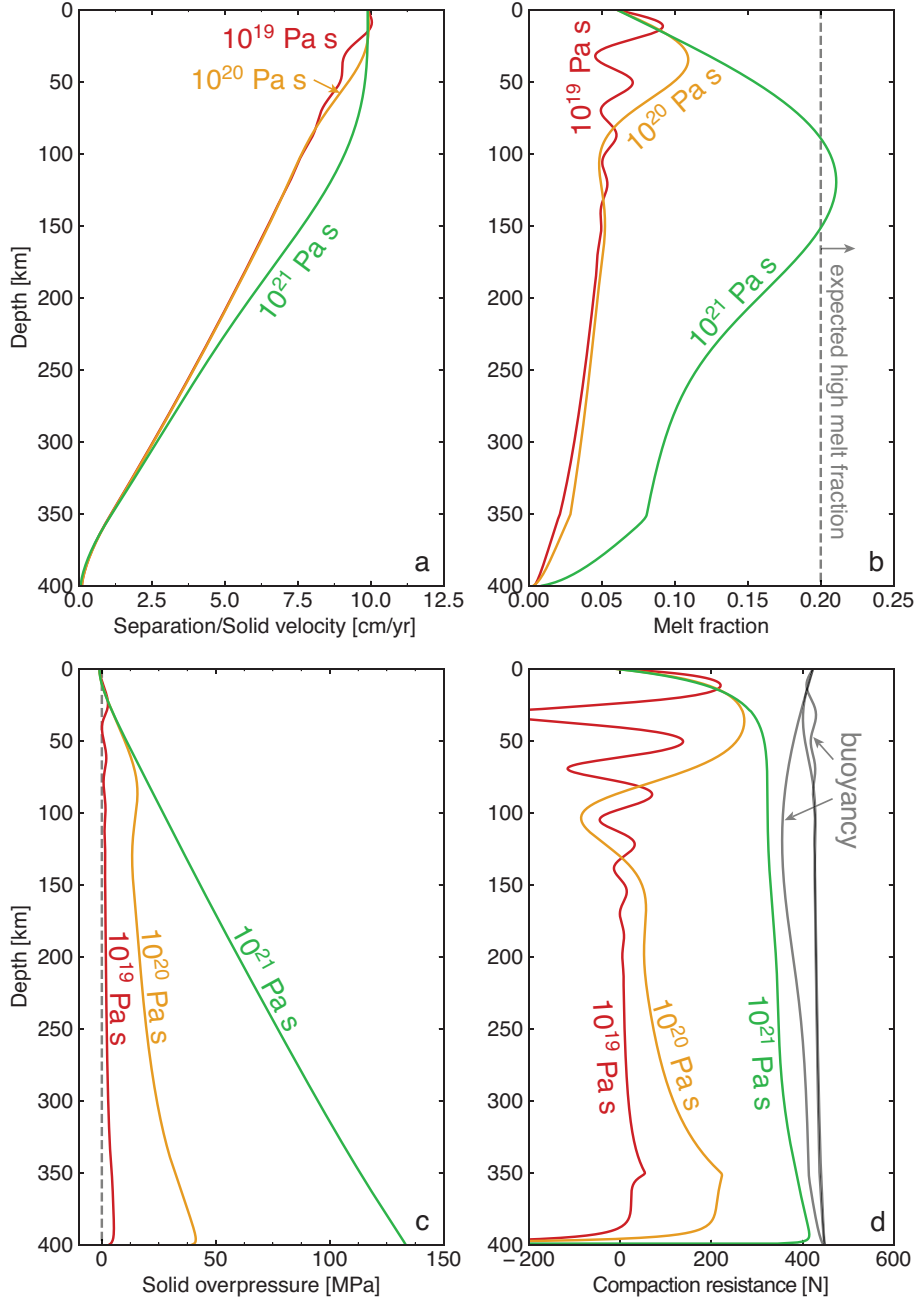}
\caption{Steady state solutions of (a) separation velocity (cm/yr), (b) melt fraction, (c) solid overpressure, and (d) forces in Equation~\ref{diffP} under $\eta_l = 10$~Pa~s and 160 TW of tidal dissipation within the layer. Colors indicate different solid viscosities $\eta_s$, corresponding to $10^{19}$ (red), $10^{20}$ (orange), and $10^{21}$~Pa~s (green). (b) The thickness of a high melt fraction layer ($\phi>0.2$) exceeds 50~km only when $\eta_s$ is larger than $10^{21}$~Pa~s. (d) Buoyancy (gray) is plotted positive upward, whereas the resistance of solid matrix to compaction (colored) are shown positive downward.}
\label{fig_solid}
\end{figure}

As discussed in Section~\ref{sec:par}, both melt and solid viscosities could vary by a few orders of magnitude with temperature and volatile content, so we explore the relationship between the porosity field and the rheological parameters under a steady state. It is noted that the velocity field is mostly unaffected by a change in the rheological parameters. The transport of latent heat dominates heat loss from the partial melt layer, and thus under the same rate of tidal dissipation, the melt flux at the top boundary ($=(1-\phi)u$) becomes similar regardless of the values of $\eta_l$ and $\eta_s$ (Figure~\ref{fig_melt}a, \ref{fig_solid}a).

When melt is less viscous, the overall degree of melting decreases because melt can escape easily under a smaller porosity (Figure~\ref{fig_melt}b). The resistance of solid matrix to compaction is insignificant except in the top boundary region, and the velocity field is determined so that viscous drag mostly balances buoyancy (Figure~\ref{fig_melt}c). The velocity profile is similar regardless of $\eta_l$, so the melt fraction decreases with a smaller $\eta_l$ to produce the same level of viscous drag. Under a tidal dissipation rate of 100~TW, a high melt fraction layer consistent with the result of \cite{Khurana2011} appears only when $\eta_l > 100$~Pa~s. 

Solid matrix with a higher bulk viscosity also results in a larger degree of melting (Figure~\ref{fig_solid}b). A larger bulk viscosity produces a greater pressure difference between the solid and melt phases (Equation~\ref{diffP}, Figure~\ref{fig_solid}c), and thus the resistance of solid matrix to compaction becomes significant at a wider depth range (Figure~\ref{fig_solid}d). In such a region characterized by compaction resistance, the profile of melt fraction exhibits oscillation. The wavelength of porosity oscillation is longer under a more viscous solid matrix (Figure~\ref{fig_solid}c), which is related to the compaction length \citep{Spiegelman1993b}:
\begin{equation}
	\delta_0 = \sqrt{\frac{k_0 \phi_0^2 \cdot \eta_s/\phi_0}{\eta_l}}.
\end{equation}
Under a similar velocity profile, the peak melt fraction increases with a longer wavelength, so a higher melt fraction is stable under a more viscous matrix. For parameters used in Figure~\ref{fig_solid}, a 50~km thick layer with $\phi > 20$~\% appears when $\eta_s$ is higher than $10^{21}$~Pa~s. The compaction length reaches $\sim$140~km in such a case, but the actual wavelength is longer than the compaction length under a larger amplitude, so the resistance to compaction plays a dominant role in governing the entire partial melt layer (Figure~\ref{fig_solid}d).

\begin{figure}[htbp]
\centering
\includegraphics[width=.9\linewidth]{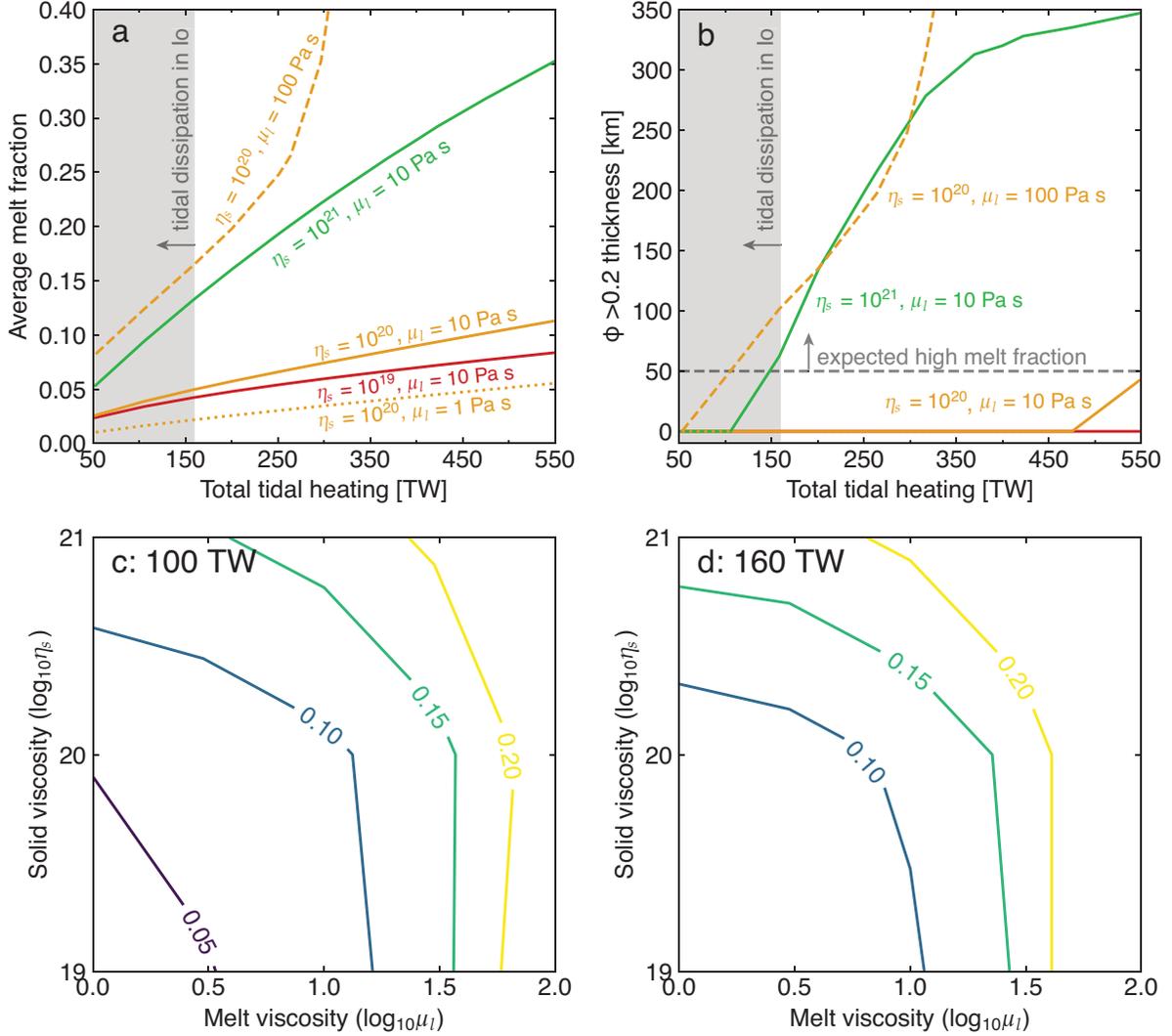}
\caption{(a) Average melt fraction and (b) the thickness of high-melt fraction layer ($\phi>0.2$) as a function of total tidal dissipation within the partial melt layer. Colors indicate different solid viscosities $\eta_s$, corresponding to $10^{19}$ (red), $10^{20}$ (orange), and $10^{21}$~Pa~s (green), and dotted, solid, and dashed lines describe melt viscosities $\eta_l$ of 1, 10, and 100~Pa~s, respectively. (c, d) Contours of the maximum melt fraction as a function of melt and solid viscosities. For a predicted rate of tidal dissipation (100-200~TW), a high melt fraction layer consistent with the observation of \cite{Khurana2011} is only stable when $\eta_l \ge 100$~Pa~s or $\eta_s \ge 10^{21}$~Pa~s.}
\label{fig_sum}
\end{figure}

We repeated the calculation for a range of values, and Figure~\ref{fig_sum} shows that a high melt fraction layer with 50~km thickness is stable only under a higher end of parameter ranges. With an estimated rate of tidal dissipation ($\sim$100~TW), a high melt fraction layer of $\phi>0.2$ can be sustained only with $\eta_l \ge 100$~Pa~s (Figure~\ref{fig_sum}a-c). Because the contribution from low-temperature heat sources may be underestimated in the observed heat flux \citep[e.g.,][]{Stevenson1988, Veeder2012}, we also tested a larger value of 160~TW, but the stability field of the high melt fraction layer did not change much (Figure~\ref{fig_sum}d). It is noted that some amount of tidal dissipation should be taking place in the overlying crust or in the underlying solid layer \citep[e.g.,][]{Peale1979, Bierson2016}, so only a fraction of the total tidal heating can be used to sustain a magmatic sponge. If $\eta_s \le 10^{20}$ and $\eta_l \le 10$~Pa~s, the rate of tidal dissipation needs to be higher than 500~TW to maintain a layer of $\phi>0.2$ (Figure~\ref{fig_sum}b), and thus a magmatic sponge is unlikely to be stable inside the present-day Io.

\subsection{Scaling law}
\begin{figure}[htbp]
\centering
\includegraphics[width=.97\linewidth]{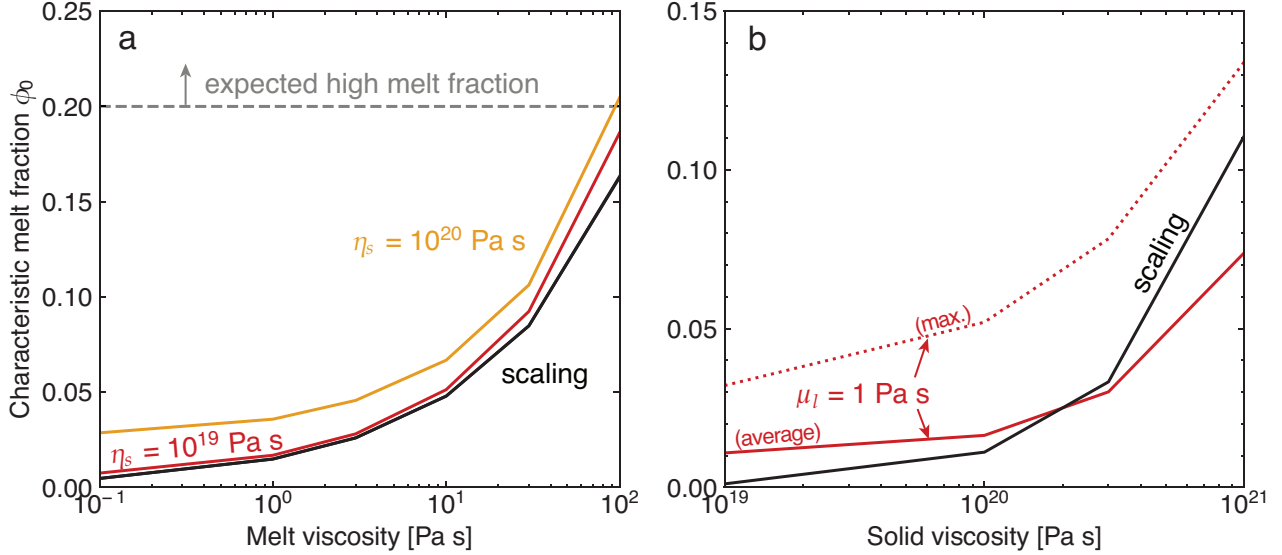}
\caption{Characteristic melt fraction of the partial melt layer as a function of rheological parameters. Black lines show values calculated using scaling laws, whereas colored ones are results from Figures~\ref{fig_ref}-\ref{fig_solid}. A surface volcanic flux of 1.5~cm/yr is assumed here, which corresponds to a surface heat flow of 100~TW. (a) Equation~(\ref{scale_melt}) is adopted as the scaling relation, whereas cases for $\eta_s = 10^{19}$ (red) and $10^{20}$~Pa~s (orange) are used to represent the calculated porosity profile.  The average melt fraction at the top 20\% of the partial melt layer is adopted as characteristic values. (b) The scaling of Equation~(\ref{scale_solid}) is shown, and a case for $\eta_l = 1$~Pa~s represents the modeled porosity profile. It is noted that Equation~(\ref{scale_solid}) predicts the average melt fraction of the partial melt layer, and the maximum melt fraction (dotted) is higher than the value indicated by the scaling law.}
\label{fig_scale}
\end{figure}

Although the details of the porosity profile depends on the imposed boundary condition, the characteristic melt fraction within the partial melt layer can be deduced from a simple scaling that follows Equation~(\ref{eom2}). In the limit where the viscosity of the solid matrix is negligible, separation flux $u$ asymptotes to
\begin{equation}
	u = \frac{k_m}{\eta_l} (1-\phi) \Delta \rho g,
\end{equation}
which is valid under $\eta_s \le 10^{20}$~Pa~s in Figure~\ref{fig_solid}. A typical melt fraction in the upper region of the partial melt layer, $\phi_0$, can then be approximated by
\begin{equation} \label{scale_melt}
	\phi_0^2 (1-\phi_0) \simeq \phi_0^2 = \frac{ \eta_l u_0}{k_0 \Delta \rho g},
\end{equation}
using the flux leaving the partial melt layer from the top boundary, $u_0$. The surface volcanic flux is estimated as $\sim$1.5~cm/yr from the global heat flux (Section~\ref{sec:litho}), but considering the amount of melt solidifying before eruption, the amount of melt leaving the partial melt layer, $(1-\phi_0) u_0$, is likely to be $\sim$7~cm/yr (Section~\ref{sec:litho}). Figure~\ref{fig_scale}a and Equation~(\ref{scale_melt}) show that the value of $\phi_0$ roughly matches the melt fraction obtained by solving Equations~(\ref{eom2}) and (\ref{cene2}).

Similarly, in the limit where the bulk viscosity of the solid matrix dominates the system, the following scaling can be obtained by integrating Equation~(\ref{eom2}):
\begin{equation}
	u \simeq \frac{\Delta \rho g (D/2)^2}{\eta_s/\overline{\phi}},
\end{equation}
where $D$ is the thickness of the partial melt layer. This scaling relates $u$ to the average melt fraction over $D$, $\overline{\phi}$, and the average melt fraction $\overline{\phi}$ can be approximated by
\begin{equation} \label{scale_solid}
	\overline{\phi} \simeq \frac{4 u_0 \eta_s}{\Delta \rho g D^2},
\end{equation}
which explains the porosity profile calculated in Figure~\ref{fig_solid} for a high solid viscosity (Figure~\ref{fig_scale}b). Therefore, even without solving for differential equations, scaling relations indicate that a layer with high melt fraction is incompatible with an observed volcanic flux on Io unless melt or solid viscosity is high. 

\subsection{Dependency on model setting}
\begin{figure}[htbp]
\centering
\includegraphics[width=.7\linewidth]{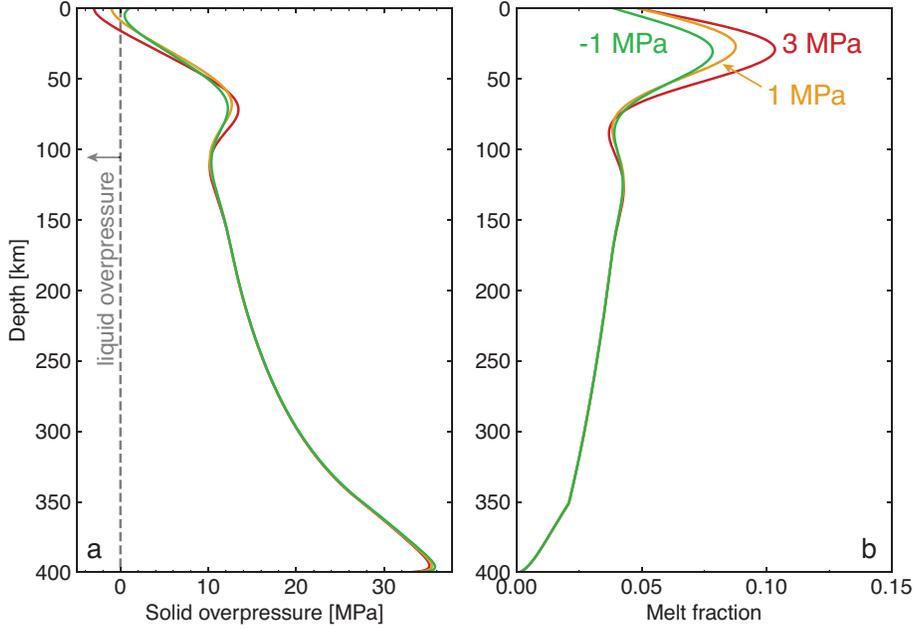}
\caption{The profiles of (a) solid overpressure (MPa) and (b) melt fraction for a case with $\eta_l = 10$ and $\eta_s = 10^{20}$~Pa~s. Colors indicate pressure difference imposed at the top boundary, corresponding to liquid overpressures of 3 (red), 1 (orange), and -1 MPa (green). The profile of melt fraction changes little with boundary conditions.}
\label{fig_tau}
\end{figure}

The most unconstrained parameter used in this study may be the pressure difference imposed at the top boundary ($\tau_0$). This value, however, has little influence on the overall results. Increasing the value of liquid overpressure by a factor of 3 increases the maximum melt fraction only by a few percent (Figure~\ref{fig_tau}). On the other hand, when a solid overpressure is set at the top boundary, the maximum value of melt fraction decreases, and a high melt fraction layer becomes less stable. Therefore, the conclusion that a magmatic sponge is unstable in Io  remains valid regardless of $\tau_0$. Liquid overpressure larger than 15~MPa has been suggested to create a high melt fraction layer beneath the crust \citep{Spencer2020a}, but the value of $\tau_0$ is unlikely to deviate largely from 0~MPa (Section~\ref{sec:boundary}). Also, partial melts can evolve under much lower melt-solid pressure differences \citep[e.g.,][]{Sleep1988, Stevenson1989}, so such a large overpressure may not be sustained in the near-surface region. 

\begin{figure}[htbp]
\centering
\includegraphics[width=.7\linewidth]{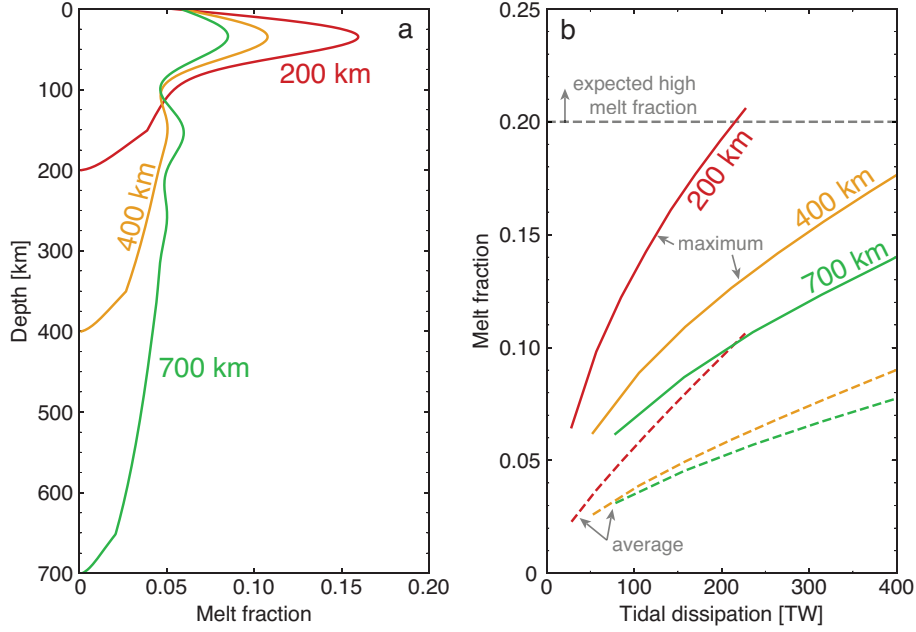}
\caption{(a) The profile of melt fraction under the tidal dissipation rate of 160~TW. Three different layer thicknesses are assumed: 200 (red), 400 (orange), and 700~km (green), and the degree of melt increases when tidal dissipation is concentrated within a narrow depth range. (b) Maximum (solid) and average (dashed) melt fraction. Results are plotted as a function of total tidal dissipation within a partially molten layer. }
\label{fig_depth}
\end{figure}

Although we assumed that tidal heating is uniformly distributed within the partial melt layer, the volumetric heating rate may be enhanced at certain depths \citep{Tyler2015, Bierson2016}. To discuss how a non-uniform distribution may affect the results, we solved for the porosity profile when tidal dissipation is concentrated in a narrower depth range. Figure~\ref{fig_depth} shows a case when 160~TW of tidal heating is applied to a 200~km-thick partially molten layer, which is half of the 400~km assumed in Figures~\ref{fig_ref}-\ref{fig_scale}. With the same total amount of dissipation, the degree of melting increases when heating is concentrated, and such a trend can be understood from the scaling law (Equations~(\ref{scale_melt}) and (\ref{scale_solid})): under the same $u_0$, $\phi$ increases with a smaller $D$. The maximum melt fraction, however, does not exceed 0.2 even when 160~TW of dissipation is applied to a 200~km-thick layer, and a high-melt fraction layer compatible with \cite{Khurana2011} requires even more concentrated heating: 160~TW of heating within a 100~km-thick layer. Considering that some fraction of tidal dissipation is happening in the solid lower mantle \citep{Bierson2016}, such a condition is unlikely to be met. For Io, a magmatic sponge is thus expected to be unstable under the current dissipation rate regardless of its distribution within the interior.


\section{Discussion}
\subsection{The fate of melt leaving the high-melt fraction layer} \label{sec:litho}
A fraction of magma leaving the high-melt fraction layer solidifies before reaching the surface \citep{Spencer2020a}, and the degree of solidification is discussed in this section. Regardless of the nature of a high-melt fraction layer, whether it is a magmatic sponge or a magma ocean, heat transport within the crust would predominantly be achieved through heat pipes \citep{OReilly1981}. Therefore, the energy balance of the near surface region can be expressed as 
\begin{equation} \label{ceneb}
	\left. -k\frac{dT}{dz}\right|_{z=z_s} + \rho c_p \Delta T v_s = \left. -k\frac{dT}{dz}\right|_{z=z_b} + \rho L \Delta v,
\end{equation}
where $\Delta T$ is a temperature difference between the erupting magma and the surface and $v_s (>0)$ is the subsiding velocity of the crust, which should be the same as the surface magma flux (Figure~\ref{fig_cartoon}c). Subscript $s$ denotes the surface, and $b$ describes the base of the layer, which is equivalent to the top of the partial melt layer in Figure~\ref{fig_ref}-\ref{fig_solid}. Conductive heat flux from the partially molten layer is more than two orders of magnitude smaller than the latent heat flux, and its magnitude decreases further towards the surface as the temperature gradient asymptotes to zero \citep{OReilly1981}. The amount of melt solidifying within this region, $\Delta v \equiv v_b - v_s$, can thus be estimated as
\begin{equation} \label{eq:deltav}
	\Delta v = v_s \frac{c_p \Delta T}{L} \simeq 3.5 v_s.
\end{equation}
With a temperature difference of $\Delta T = 1400$~K, this equation suggests that nearly 3.5 times the surface magma flux solidifies before magma erupts to the surface \citep{Spencer2020a}. The latent heat of solidifying magma is used to heat the cold subsiding crust. 

It is noted that conductive heat flux at the surface could still be significant in some regions, even though the temperature gradient decreases toward the surface. Igneous events can leave residual heat at erupted sites \citep{Stevenson1988} or create a magma chamber beneath the surface \citep{Davies2006_Io}. Some fraction of heat transported through heat pipes would create low-temperature thermal sources, and it may not be included in the currently observed heat from volcanic hot spots \citep[$\sim$62~TW;][]{Veeder2012}. Further missions are warranted to understand its magnitude.

The value of $v_s$ can be estimated from the energy balance of Io. At steady state, the total tidal heating should balance surface heat flux:
\begin{equation} \label{eq:v}
	\int_{V_m} \psi dV \simeq 4 \pi r_s^2 \cdot \rho \left( L + c_p \Delta T \right) v_s,
\end{equation}
where $V_m$ is the volume of partial melt layer experiencing tidal deformation. Assuming a total dissipation rate of $122\pm40$~TW \citep{Veeder2004}, Equation~(\ref{eq:v}) suggests that the globally-averaged surface volcanic flux is $v_s \simeq 1.5\pm 0.5$~cm/yr, and the melt flux leaving the partial melt layer would be $v_b \simeq 6.8\pm 2.2$~cm/yr.

\subsection{Characterization of the subsurface structure} \label{sec:transit}
The solidification of upwelling melt ($\Delta v$ in Section~\ref{sec:litho}) is considered to be happening in a ``transition'' zone, which connects the crust and the partial melt layer discussed in Section~\ref{sec:res}. The top boundary considered in Section~\ref{sec:res} corresponds to the base of this transition zone. The transition zone is cooled from above by the cold subsiding crust, while upwelling melt delivers heat from the deeper region. The heat balance of the boundary layer is thus characterized as
\begin{equation}
  	\rho L \nabla \cdot \kak{ (1-\phi) u} \simeq k \nabla^2 T (<0),
\end{equation}
which describes the conversion of latent heat into conductive heat. The thickness of this boundary layer, $D$, can be estimated using scaling analysis ($\rho L \Delta U \sim k \Delta T/D$). The upwelling flux that solidifies before eruption is denoted as $\Delta U$, and $\Delta U$ of 5 cm/yr is expected for Io from Equation~(\ref{eq:deltav}). The scaling suggests that $\Delta T/D \sim0.5$~K~m$^{-1}$, so the layer thickness should only be $\sim$1~km when the temperature change across this boundary layer is 500~K. The exact value of $D$ requires a detailed understanding of how melt migrates and solidifies within the crust and the boundary layer, but this rough scaling shows that the boundary layer is considerably thinner than the radius of Io \citep{Spencer2020a}. 

The temperature is expected to rapidly drop in a narrow depth range, and with melt fraction quickly decreasing as well, the standard recipe of percolation would cease to apply in the transition zone. How a percolative flow induces cracking and turns into a channelized flow remains elusive and is left for future work. For now, our calculation is based on a fact that the value of liquid overpressure would be around $\tau_0 \sim 0$~MPa at the top boundary, considering that melt flux ($u$) starts to decrease towards the surface in the transition zone (Section~\ref{sec:boundary}). This approach allows us to solve for the thermal state of the crust and partial melt layer separately, and calculations in Section~\ref{sec:res} would not depend on the details of the transition zone.

\subsection{The interior of a partially molten body}
Figure~\ref{fig_sum}a shows that the amount of tidal heating required to sustain a partial melt layer increases with the average melt fraction, whereas the rate of tidal dissipation is known to peak at a low melt fraction \citep[$\phi \sim$0.05-0.1;][]{Ross1986, Zahnle2015}. A planetary body with a partially molten interior is thus likely to maintain an equilibrium state with an average melt fraction of $\phi \sim$0.1. When $\phi \gg 0.1$, heat loss by percolation is expected to be larger than that the tidal dissipation, and the average melt fraction is expected to decrease with time. An excess amount of melt would either escape to the surface or separate from the solid matrix to create a subsurface magma ocean. A rheological model would be needed to solve for the detailed structure, but the typical melt fraction of a partially molten layer experiencing extreme tidal heating is likely to fall into the range of $\phi \sim 0.05-0.1$. 

\section{Conclusions}
In this study, we calculated the amount of internal heating necessary to sustain a magmatic sponge: a high degree of melting with interconnected solid. Our results suggest that the presence of such a partially molten layer requires internal heating much larger than the rate of tidal dissipation expected within Io. A high partial melt layer could be stable when the rheological parameter is in the higher end of the estimated range, but such a case is less likely given that Io contains an abundant amount of sulfur in the interior. Because the amount of internal heating is insufficient to maintain a high degree of melting, a magmatic sponge would separate into two phases, creating a subsurface magma ocean. We note that the magma ocean may not necessarily be pure liquid but could contain some amount of crystals. Crystals start to disaggregate once the melt fraction reaches $\phi \ge 0.4$, and a partially molten material would behave rheologically as liquid as long as $\phi \ge 0.4$ \citep[e.g.,][]{Abe1993a, Solomatov1993a}. A crystal-rich magma ocean in Io has been supported by petrologic arguments \citep{Keszthelyi1999}.

Governing equations of such a subsurface magma ocean would differ from Equations~(\ref{eom2}, \ref{cene2}). Some tidal dissipation may occur in this fluid layer \citep{Tyler2015}, but even without such heating, sufficient heating could be provided from the underlying partial melt layer (Figure~\ref{fig_cartoon}). A magma ocean may simply be acting as a layer that transports heat from the partial melt layer to the crust.  Compositions of such a magma ocean and the partial melt layer are likely to be different, but knowledge of petrology would be required to delineate its detail.

A strong induction signal from the Io's interior has recently been questioned \citep{Blocker2018}, but if a melt-rich layer is indeed present in the subsurface of Io, such a layer should exist as a magma ocean. Its existence may be confirmed if a Love number around $\sim$0.5 is measured by Juno flybys. If the measurement otherwise suggests a larger value around $\sim$0.1, a strong induction signal may simply be a result of a plasma interaction in Io's atmosphere \citep{Blocker2018}. In such a case, the melt fraction is likely to be smaller than $\phi<0.2$ in the Io's interior, and a high-melt fraction layer would be absent within Io.

\acknowledgments This work was motivated by NASA's Juno mission. Y.M. was supported by Stanback Postdoctoral Fellowship from Caltech Center for Comparative Planetary Evolution. 

\appendix
\section{Method Details} 
\subsection{Non-dimesionalized equations} \label{sec:nond}
Using scales $z = \delta_0 z^*$, $u = u_0 u^*$, and $\phi = \phi_0 \phi^*$, Equation~(\ref{eom2}) is non-dimensionalized as follows:
\begin{equation}
	\frac{k_0 \Delta \rho g}{\eta_l u_0} \kak{1 - \phi_0 \phi^*} + \frac{k_0 \zeta_0}{\eta_l \delta_0^2} \frac{d}{dz^*} \kak{\frac{1 - \phi_0 \phi^*}{\phi^*} \frac{du^*}{dz^*}} = \frac{u^*}{\phi_*^n},
\end{equation}
where $\zeta_0 = \eta_s/\phi_0$ is a reference bulk viscosity at $\phi = \phi_0$. By taking $\delta_0$ as the compaction length (Equation~\ref{dcomp}) and $u_0$ as $(k_0 \Delta \rho g)/\eta_l$, the equation for momentum conservation can be simplified to \citep{Scott1984}
\begin{equation} \label{neom}
	 \kak{1 - \phi_0 \phi^*} + \frac{d}{dz^*} \kak{\frac{1 - \phi_0 \phi^*}{\phi^*} \frac{du^*}{dz^*}} = \frac{u^*}{\phi_*^n}.
\end{equation}
The conservation of energy is also non-dimensionalized as 
\begin{equation} \label{neoc}
  	\nabla_* \cdot \kak{ (1-\phi_0 \phi^*) u^*}  =  \frac{\kappa}{\delta_0 u_0} \frac{c_p T_0}{L} \nabla_*^2 T + \frac{\delta_0}{u_0} \frac{\psi}{\rho L},
\end{equation}
and these two equations are solved in our model.


\end{document}